\documentclass[3p,times,twocolumn]{elsarticle}
\pdfoutput=1 
\usepackage{ecrc}
\usepackage{natbib}

\volume{00}

\firstpage{1}

\journalname{Nuclear Physics B Proceedings Supplement}

\runauth{R. Aloisio et al.}

\jid{nuphbp}

\jnltitlelogo{Nuclear Physics B Proceedings Supplement}

\usepackage{amssymb}
\usepackage[figuresright]{rotating}

\begin{document}

\begin{frontmatter}

\dochead{}

\title{Propagation of Ultra High Energy Cosmic Rays and \\the Production of Cosmogenic Neutrinos}

\author[gssi,arcetri]{R. Aloisio} 
\author[lngs]{D. Boncioli} 
\author[aq]{A. Di Matteo} 
\author[lngs]{A. Grillo} 
\author[aq,gssi]{S. Petrera}
\author[ipno]{F. Salamida}

\address[gssi]{Gran Sasso Sicence Insitute (INFN), L'Aquila, Italy}
\address[arcetri]{INAF - Arcetri Astrophysical Observatory, Florence, Italy}
\address[lngs]{INFN - Laboratori Nazionali del Gran Sasso, Assergi, Italy}
\address[aq]{INFN and Department of Physical and Chemical Sciences, University of L'Aquila, Italy}
\address[ipno]{Institut de Physique Nucleaire d'Orsay (IPNO), Universite Paris 11, CNRS-IN2P3, France\\
(currently at the Istituto Nazionale Fisica Nucleare, Milano Bicocca, Italy) }

\begin{abstract}
We present an updated version of the {\it SimProp} Monte Carlo code to study the propagation of ultra high energy cosmic rays in astrophysical backgrounds computing the cosmogenic neutrino fluxes expected on earth. The study of secondary neutrinos provides a powerful tool to constrain the source models of these extremely energetic particles. We will show how the newly detected IceCube neutrino events at PeV energies together with the the latest experimental results of the Pierre Auger Observatory and Telescope Array experiment are almost at the level of excluding several hypothesis on the astrophysical sources of ultra high energy cosmic rays. Results presented here can be also used to evaluate the discovery capabilities of future high energy cosmic rays and neutrino detectors. 
\end{abstract}


\end{frontmatter}

\section{Introduction}
\label{intro}

The study of Ultra High Energy Cosmic Rays (UHECR) started already in '60 with the first pioneering observations of the Volcano Ranch experiment, that, in 1962, observed the first cosmic ray event with energy larger than $10^{20}$ eV \cite{Linsley:1963km}. Nowadays the most evolute experiments observing UHECR are the Pierre Auger Observatory in Argentina \cite{ThePierreAuger:2013eja}, far the largest experimental setup devoted to the study of these particles, and the Telescope Array (TA) experiment \cite{Tinyakov:2014lla}, placed in the United States, with roughly 1/10 of the Auger statistics at the highest energies.  

The experimental study of UHECR clarified few important characteristics of these particles: (i) UHECR are charged particles with a limit on photon and neutrino fluxes around $10^{19}$ eV at the level of few percent and well below respectively \cite{Abraham:2009qb,Abu-Zayyad:2013dii,Abreu:2013zbq}, (ii) the spectra observed on earth show a slight flattening at energies around $5\times 10^{18}$ eV (called the ankle) with (iii) a steep suppression at the highest energies $\simeq 10^{20}$ eV \cite{ThePierreAuger:2013eja,Abu-Zayyad:2013qwa}.

The propagation of UHECR from the source to the observer is conditioned by the expansion of the universe and the interaction with astrophysical backgrounds, namely the Cosmic Microwave Background (CMB) and the Extragalactic Background Light (EBL). While the propagation of nucleons\footnote{Hereafter discussing freely propagating UHE nucleons we will always refer only to protons because the decay time of neutrons is much shorter than all other time scales involved \cite{Aloisio:2008pp,Aloisio:2010he}.} is conditioned only by the CMB field the propagation of nuclei is also affected by the EBL. Apart from the expansion of the universe that, adiabatically, reduces the energy of any propagating particle, the processes that involve protons are: (i) pair production and (ii) photo-pion production; those involving nuclei are: (i) pair production and (ii) photo-disintegration \cite{Aloisio:2008pp,Aloisio:2010he}. 

One of the most important observables in the physics of UHECR is certainly the chemical composition of these particles. The experimental observations of the composition are not conclusive, with different results claimed by TA and Auger. The analysis performed by TA is compatible with a proton dominated composition at all energies, starting from the lowest around $10^{18}$ eV up to the highest at $10^{20}$ eV \cite{Tinyakov:2014lla} . On the other hand the observations of Auger show a more rich phenomenology with the lowest energies dominated by protons and, starting from energies around $3\times 10^{18}$ eV, a composition more and more dominated by heavier nuclei with a strongly reduced number of protons at energies above $2\times 10^{19}$ eV \cite{ThePierreAuger:2013eja}. 

Restricting the analysis to a pure proton composition, as appropriate to the interpretation of TA data, the UHECR observations can be elegantly explained by the dip model \cite{Berezinsky:2002nc,Aloisio:2006wv}. In the framework of this model the flux behaviour in the region of the ankle is due to the effect of the proton pair-production process on the CMB radiation field  \cite{Berezinsky:2002nc}. While the strong flux suppression at the highest energies is the effect of the photo-pion production process, the so-called Greisen-Zatsepin-Kuzmin (GZK) cut-off \cite{Greisen:1966jv,Zatsepin:1966jv}. Taking into account the sole Auger data, to reach a reasonable agreement with observations, one should consider mixed compositions. Assuming that protons give their principal contribution only at the lowest energies $\le 3\times 10^{19}$ eV \cite{Aloisio:2013hya,Taylor:2013gga}, below the photo-pion production threshold $\sim 6\times 10^{19}$ eV. In this case the flux suppression observed at the highest energies is due to the photo-disintegration process suffered by heavy nuclei \cite{Aloisio:2013hya,Aloisio:2009sj}.

The production of secondary particles due to the interactions of UHECR with background photons is strongly dependent on the chemical composition. As was first realised in \cite{Beresinsky:1969qj}, the proton content at the highest energies is a crucial quantity that regulates the fluxes of secondary (cosmogenic) photons and neutrinos. In the present paper, implementing an updated version of the {\it SimProp}\footnote{{\it SimProp} is available upon request writing to the authors or at \it{SimProp-dev@aquila.infn.it}.} Monte Carlo (MC) code \cite{Aloisio:2012wj}, we will discuss the production of secondary neutrinos by the propagation of UHECR. The study presented here, discussed in detail in \cite{Aloisio:2015ega}, has a twofold interest: from one side using the latest observations of the IceCube \cite{Aartsen:2013jdh} and Auger  \cite{Abreu:2013zbq} we can already draw interesting conclusions on the sources of UHECR; on the other side this study should be intended as a benchmark computation to asses the discovery capabilities of the next generation experiments.

\section{UHECR models and secondary neutrinos}
\label{sec:scen}

There are two processes by which neutrinos can be produced in the propagation of UHECRs: (i) the decay of charged pions produced by photo-pion production, $\pi^{\pm}\to \mu^{\pm} + \nu_\mu(\bar{\nu}_\mu)$, and the subsequent muon decay $\mu^{\pm}\to e^{\pm}+\bar{\nu}_\mu(\nu_\mu)+\nu_e(\bar{\nu}_e)$; (ii) the beta decay of neutrons and nuclei produced by photo-disintegration: $n \to p + e^{-} + \bar{\nu}_e$, $(A,Z) \to (A,Z-1) + e^{+} + \nu_e$, or $(A,Z) \to (A,Z+1) + e^{-} + \bar{\nu}_e$. These processes produce neutrinos in different energy ranges: in the former the energy of each neutrino is around a few per cent that of the parent nucleon, whereas in the latter it is less than one part per thousand (in the case of neutron decay, larger for certain unstable nuclei). This means that in the interactions with CMB photons, with a Lorentz factor threshold around $\Gamma\gtrsim 10^{10}$, neutrinos are produced with energies of the order of $10^{18}$ eV and $10^{16}$ eV respectively. Interactions with EBL photons contribute with a much lower probability respect to CMB photons, affecting a small fraction of the propagating protons and nuclei. Neutrinos produced through interactions with EBL, characterised by lower thresholds, have energies of the order of $10^{14}$ eV in the case of photo-pion production and $10^{16}$ eV in the case of neutron decay.   

Neutrinos produced by the interaction of UHECR, because of their extremely low interaction rate, arrive on Earth unmodified with the overall universe contributing to their flux. This is an important point that makes neutrinos a viable probe not only of the chemical composition of UHECR but also of the cosmological evolution of sources that, as we will show below,  can be also constrained by the neutrino flux observed on Earth.

In the following we will consider the two cases of the dip model and mixed composition, discussing the expected neutrino flux with different assumptions on the cosmological evolution of sources. We will consider the case of sources (see \cite{Aloisio:2015ega} and reference therein) (i) with no cosmological evolution, (ii) with the same cosmological evolution of Active Galactic Nuclei (AGN), supposed to play a role in particles acceleration till the highest energies \cite{Berezinsky:2002nc}, and (iii) with the cosmological evolution of the Star Formation Rate (SFR).  All computations presented here, discussed in detail in \cite{Aloisio:2015ega}, are performed under the assumption of a homogenous distribution of sources. This assumption does not affect the expected neutrino spectra because in the case of neutrinos the overall universe, till the maximum red-shift, contributes to the fluxes. Possible flux variations due to a local inhomogeneity in sources distribution gives a negligible contribution to the total flux. We also fix a maximum red-shift of the sources $z_{max}=10$, which is the typical red-shift of the first stars (pop III) \cite{Berezinsky:2011bb}, in any case if $z_{max}>3$ the expected fluxes of primary and secondary particles is almost independent of $z_{max}$ \cite{Aloisio:2015ega}. 

Once produced at cosmological distances neutrinos travel toward the observer almost freely, the opacity of the universe to neutrinos starts to be relevant only at the highest red-shifts namely at $z>10$ \cite{Gondolo:1991rn,Weiler:1982qy}. Therefore, given the assumptions discussed above, in our computations we have neglected any effect due to neutrino propagation apart from the adiabatic energy losses due to the expansion of the universe \cite{Aloisio:2008pp,Aloisio:2010he}. 

\begin{figure}
\includegraphics[width=0.5\textwidth]{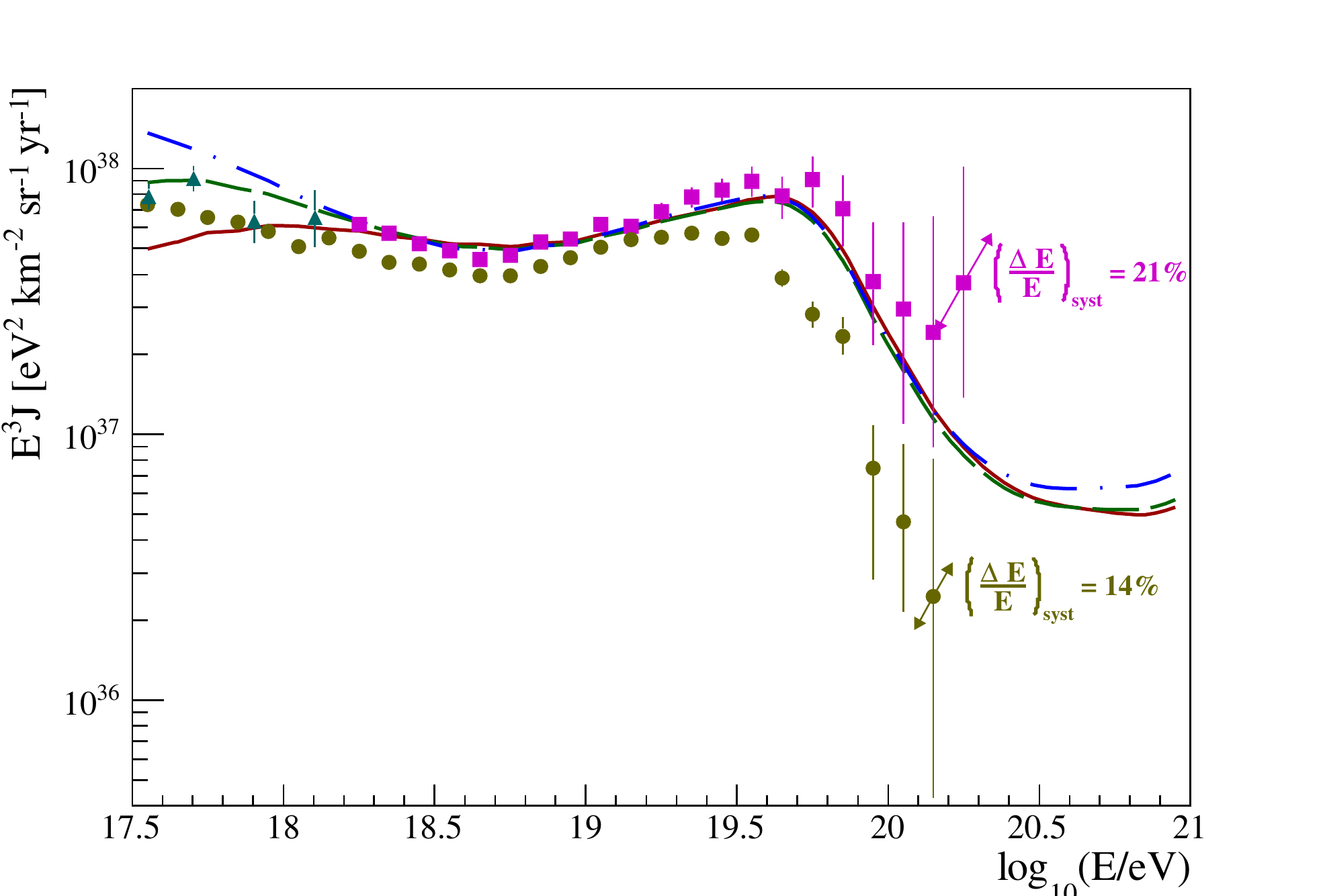} 
\caption{Fluxes of UHECR expected at Earth in the case of the dip model (pure protons). Fluxes are computed in the three cases: AGN cosmological evolution $\gamma_g=2.4$ dot-dashed line, SFR evolution with $\gamma_g=2.5$ dashed line and no cosmological evolution with $\gamma_g=2.6$ solid line. Experimental data are those of TA \cite{Tinyakov:2014lla} and Auger \cite{ThePierreAuger:2013eja}.}
\label{fig1}
\end{figure}

\begin{figure}
\includegraphics[width=0.5\textwidth]{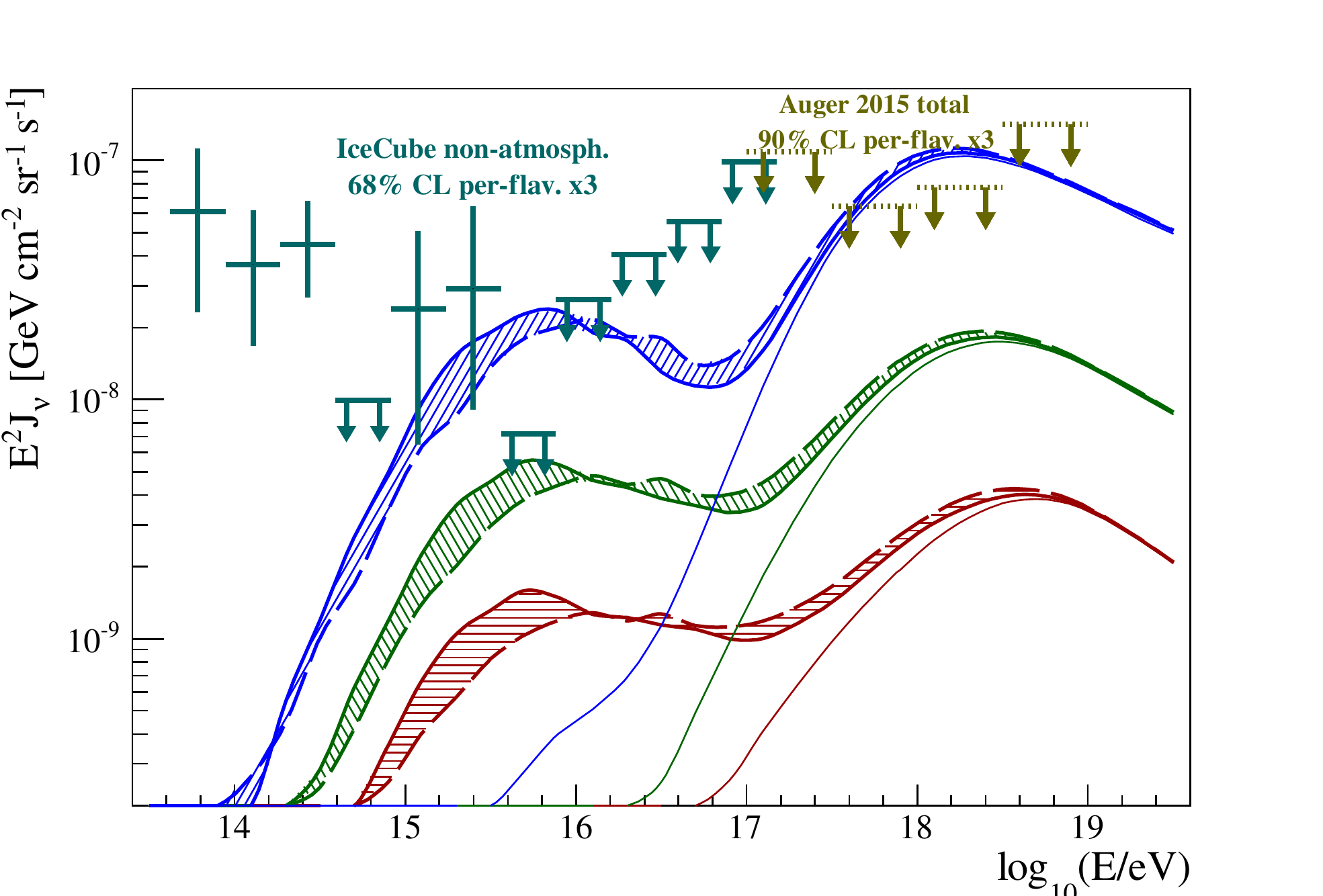}
\caption{Fluxes of neutrinos expected at Earth in the dip model with the three different assumptions on the cosmological evolution of sources, from top to bottom: AGN evolution blu band, SFR evolution green band, no evolution red band. Fluxes are plotted as coloured bands to highlight the uncertainties connected with the EBL cosmological evolution assumed (see text). Experimental data on neutrino fluxes are by IceCube \cite{Aartsen:2013jdh} and Auger \cite{Abreu:2013zbq} as labeled.} 
\label{fig2}
\end{figure}

As discussed in detail in \cite{Aloisio:2015ega}, the case of the dip model gives the highest flux of secondary neutrinos as it follows from the high efficiency of the photo-pion production process that is the major responsible for the production of secondary particles. In order to reproduce the TA observations the injection power law index $\gamma_g$ needs to be changed according to the cosmological evolution: we used $\gamma_g = 2.6$, $2.5$, and $2.4$ respectively in the three cases of no cosmological evolution, SFR and AGN. The result in terms of UHECR flux is plotted in figure \ref{fig1}, together with the TA data. In figure \ref{fig2} we plot the corresponding neutrino flux in the three cases of cosmological evolution: no evolution (red band), SFR evolution (green band) and AGN evolution (blu band). The flux of neutrinos depends on the assumptions made about the EBL cosmological evolution, therefore we have plotted the corresponding fluxes as coloured band to highlight the uncertainties related to the EBL evolution, the EBL evolution models used are those of Stecker \cite{Stecker:2005qs} and Kneiske \cite{Kneiske:2003tx}.

Results presented in figure \ref{fig2} are quite interesting, showing that models with cosmological evolution of sources stronger than the case of AGN predict a total neutrino flux (blu band in figure \ref{fig2}) almost at the level of the experimental limits fixed by Auger and the observations of IceCube. Moreover, cosmogenic neutrinos can explain the IceCube events only at PeV energies not below. 

Let us now consider the case of a mixed composition, as appropriate to the interpretation of the Auger data. As shown by \cite{Aloisio:2013hya} scenarios with mixed composition have peculiar characteristics that, in order to describe Auger observations, imply different source families. At the highest energies ($E>5\times 10^{18}$ eV) Auger observes predominantly heavy elements, with $A \gtrsim 12$, while at the lowest energies, as a matter of fact, all UHECR experiments observe a light composition dominated by protons. To reproduce such observations two classes of sources are needed: one that injects only light elements with low maximum energy and steep injection $\gamma_g>2.3$ and one that injects also heavy elements with higher maximum energy and an extremely flat injection $\gamma_g=1$ \cite{Aloisio:2013hya}.
\begin{figure}
\includegraphics[width=0.5\textwidth]{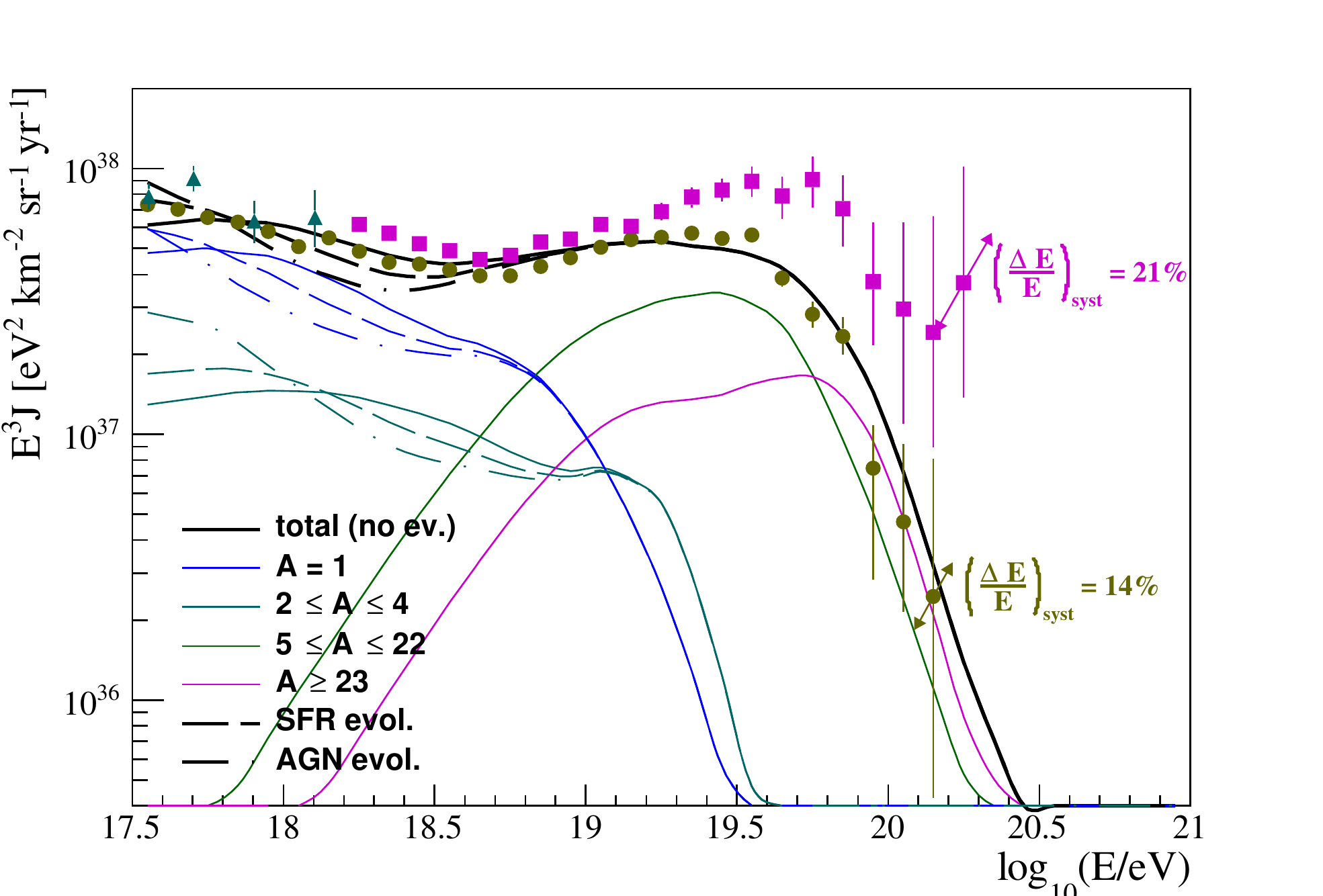} 
\caption{Fluxes of UHECR expected at Earth in the case of the mixed composition model of \cite{Aloisio:2013hya}. Experimental data are those of Auger \cite{ThePierreAuger:2013eja} and TA \cite{Tinyakov:2014lla}. Cosmological evolution of sources is taken into account only in the case of light component with: AGN cosmological evolution $\gamma_g=2.5$ dot-dashed lines, SFR evolution with $\gamma_g=2.6$ dashed lines and no evolution with $\gamma_g=2.7$ solid lines.}
\label{fig3}
\end{figure}
In the case of nuclei cosmological evolution severely constrains sources injecting heavy elements. In this case the effect of photo-disintegration of nuclei produces too many secondary protons already assuming a cosmological evolution as for the SFR and AGN cases. Therefore, in this case, we have considered only cosmological evolution for sources providing the light component (p and He) leaving without evolution those sources providing heavy components. 
Results are presented in figure \ref{fig3}, where we plot UHECR fluxes together with Auger data, and in figure \ref{fig4}, where we plot the corresponding neutrino fluxes. Light elements are injected with a power law index $\gamma_g=2.7, 2.6, 2.5$ in the three cases of no evolution, SFR and AGN evolutions respectively. As discussed in \cite{Aloisio:2013hya,Aloisio:2015ega}, this combination of sources and injections reproduces fairly well Auger observations on both flux (see figure \ref{fig3}) and chemical composition \cite{Aloisio:2013hya,Aloisio:2015ega}, giving a neutrino flux which is substantially below the Auger observations and almost at the level of the IceCube observations at PeV energies in the case of AGN cosmological evolution and well below in the other cases of cosmological evolution.

\section{Conclusions} 
\label{sec:res}

We conclude stressing the importance of the observation of secondary neutrinos produced by the propagation of UHECR, as discussed in \cite{Aloisio:2015ega} the observation of secondary neutrinos can be extremely important in order to characterise the possible sources of UHECR. As follows from figure \ref{fig2} in the case of the dip model, assuming the AGN cosmological evolution, the neutrino flux is almost at the level of being detected by both Auger and IceCube. In the case of source models with acceleration of heavy elements, the production of secondary neutrinos is quite suppressed respect to the case of pure protons, well below the detection thresholds of Auger and IceCube, this makes the non-observation of secondary neutrinos an indirect confirmation of the picture emerging from Auger data. 

\begin{figure}
\includegraphics[width=0.5\textwidth]{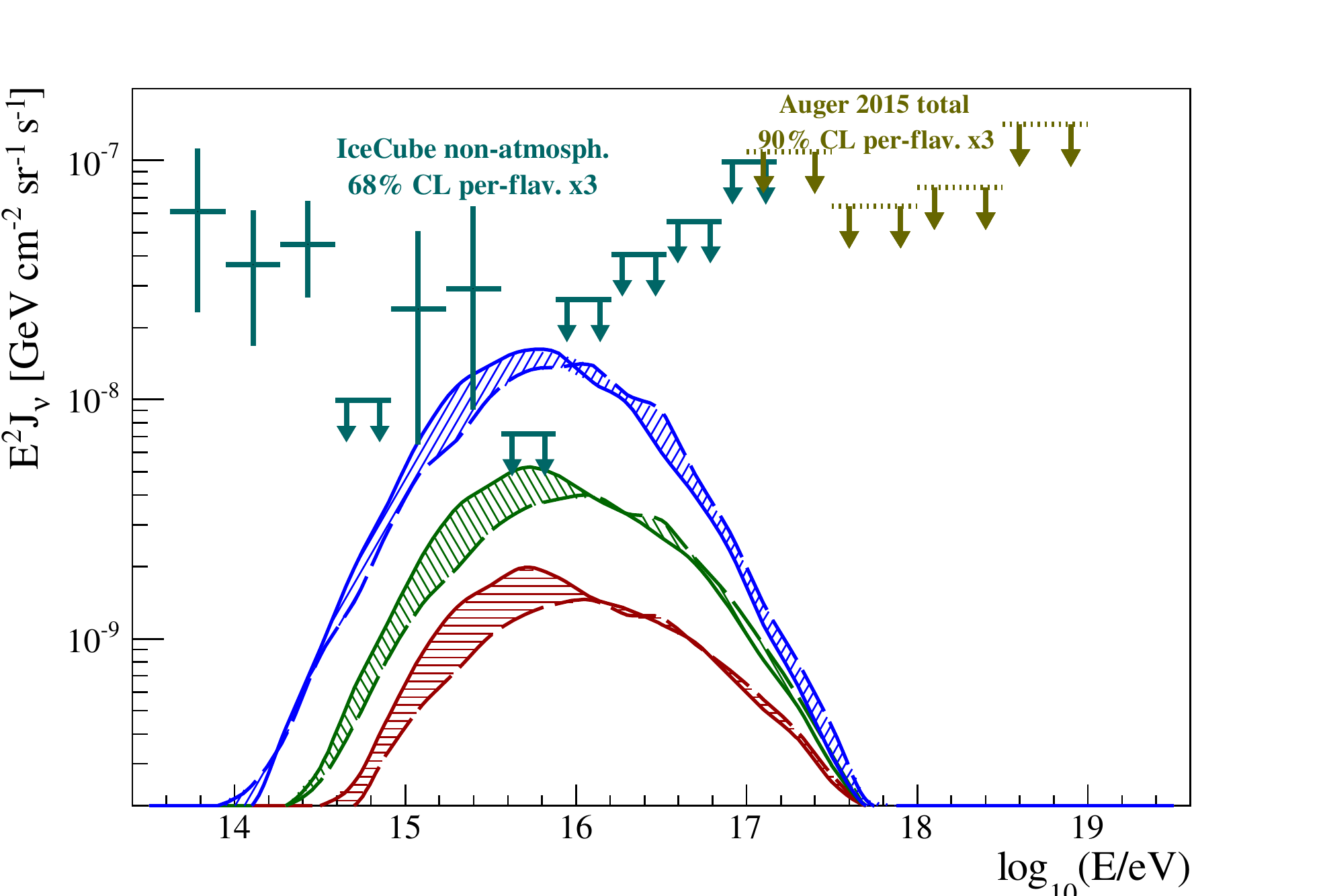} 
\caption{Fluxes of neutrinos expected at Earth in the case of the mixed composition model of \cite{Aloisio:2013hya}. The coloured bands refers to the three cases of cosmological evolution as discussed in the text, with the same color codes of figure \ref{fig2}.}
\label{fig4}
\end{figure}

\nocite{*}
\bibliographystyle{elsarticle-num}
\bibliography{ref_short}

\end{document}